\documentclass[manuscript]{raa}            
\usepackage{graphicx,times}
\usepackage{natbib}

\usepackage{deluxetable}

\providecommand{\bm}[1]{\mbox{\boldmath$#1$\unboldmath}}

\begin{document}

   \title{Investigation of Star Formation toward the Sharpless~155 H~II region
}

 \volnopage{ {\bf 20xx} Vol.\ {\bf 9} No. {\bf XX}, 000--000}
   \setcounter{page}{1}

   \author{Ya Fang Huang
      \inst{1}
   \and Jin Zeng Li
      \inst{1}
   \and Travis A. Rector
     \inst{2}
\and Zhou Fan
     \inst{1}}


   \institute{National Astronomical Observatories, Chinese Academy of Sciences,
             Beijing 100012, China; {\it huangyf@nao.cas.cn}\\
     \and
University of Alaska Anchorage, 3211 Providence Drive, Anchorage, AK 99508 USA\\
\vs \no
   {\small Received 2013 November 21; accepted 2014 April 10 }
}

\abstract{We present a comprehensive study of star formation toward the H~$\textsc{ii}$ region S155.
Star-formation activities therein were investigated based on multi-wavelength data from optical to the far-infrared. The surface density distribution of selected 2MASS sources toward S155 indicates the existence of a compact cluster, which is spatially consistent with the position of the exciting source of the H~$\textsc{ii}$ region, HD 217086. A sample of more than 200 excessive emission sources in the infrared were selected based on their 2MASS color indices. The spatial distribution of the sample sources reveals the existence of three young sub-clusters in this region, among which, sub-cluster A is spatially coincident with the bright rim of the H~$\textsc{ii}$ region. In addition, photometric data from the WISE survey were used to identify and classify young stellar objects (YSOs). To further explore the evolutionary stages of the candidate YSOs, we fit the spectral energy distribution (SEDs) of 44 sources, which led to the identification of 14 Class I, 27 Class II, and 3 Class III YSOs. The spatial distribution of the classified YSOs at different evolutionary stages presents a spatio-temporal gradient, which is consistent with a scenario of sequential star formation. On the other hand, Herschel PACS observations toward the interface between S155 and the ambient molecular cloud disclose an arc-shaped dust layer, the origin of which could be attributed to the UV dissipation from the early type stars e.g. HD 217061 in S155. Four dusty cores were revealed by the Herschel data, which hints for new generations of star formation.
}

   \authorrunning{Huang et al.}
   \titlerunning{Investigation of Star Formation toward the Sharpless~155}  
   \maketitle


%
%
\section{Introduction}           
\label{sect:intro}

Cepheus (Cep) OB3 is a very young association at a distance of about 800~pc from the Sun \citep{1993A&A...273..619M}. It covers a region from $22^h46^m00^s$ to $23^h10^m00^s$ in right ascension and from $+61^\circ$ to $+64^\circ$ in declination. It is mainly composed of two subgroups: the older, Cep OB3a, the largest projected dimension of which is about 17~pc, and the younger, Cep OB3b, more compact and closer to Cep molecular cloud \citep{2003MNRAS.341..805P}.
Cep OB3 association has always been considered to be a very good example of large-region sequential star formation according with the model of \cite{1977ApJ...214..725E}, where supernova remnants and stellar winds of an older stellar cluster compress the ambient clouds and trigger the formation of a second generation of stars \citep{1979ApJ...233..163S}.

\begin{figure}[h!!!]
   \centering
   \includegraphics[width=10cm, angle=0]{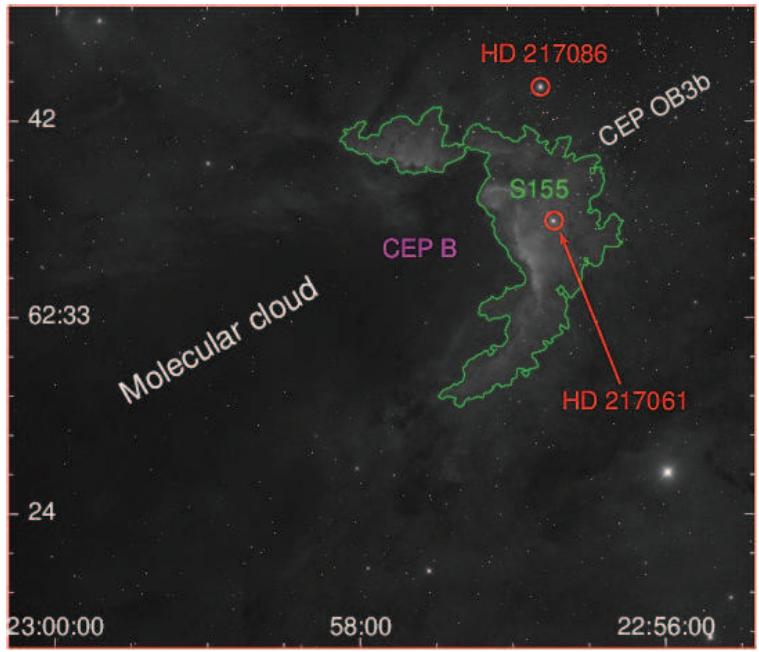}
   \begin{minipage}[]{100mm}
   \caption{H$\alpha$ narrow band image covering $30'~\times~30'$ of the S155 and its nearby region. Cep B, the hottest component of the Cep molecular cloud, is located in the center of the field. Cep OB3b, the younger subgroup of Cep OB3 association, lies to the northwest. The interface between the molecular cloud and Cep OB3b is delineated by the H~$\textsc{ii}$ region, S155. And the exiting stars HD~217086 and the illumination star HD~217061 are labeled. North is up; east is to the left.}
   \label{HIST}
   \end{minipage}
   \end{figure}

The interface between Cep OB3b and the Cep molecular cloud is clearly delineated by the optically bright H~$\textsc{ii}$ region Sharpless 155 (S155, see Figure~\ref{HIST}), where neutral material is ionized and heated by the radiation of the O7 star HD~217086 and the illumination star, HD~217061 \citep{1986PASP...98.1294L}. Both of them belong to the youngest generation of the Cep OB3b association \citep{1981A&A....98..295P}. The photodissociation region (PDR) at S155 is favorably oriented to reveal the progression of star formation. Near-infrared studies \citep{1993A&A...273..619M,1995A&A...303..881T},CO \citep{1992A&A...265..733M},far-infrared, and radio continuum \citep{1978A&A....69..199F,1995A&A...303..881T} have revealed a few young stellar objects (YSOs) embedded in the Cep cloud behind the PDR. Sources with high extinction have been detected on the edge of the Cep molecular cloud. They maybe represent a third generation of star formation triggered by the expansion of the H~$\textsc{ii}$ region. This scenario of triggered star formation has been recently strengthened by \cite{2006ApJS..163..306G, 2009ApJ...699.1454G} with the Chandra X-ray surveys and Spitzer archived data.

In this paper, we present and discuss the scenario of sequential and triggered star formation in S155 and its nearby region based on 2MASS, WISE and the Herschel PACS data. Surface density distribution of the 2MASS PSC sources presents the existence of compact clusters of pre-main sequence stars (PMS). The WISE photometric data help to remove contaminants and fit spectral energy distributions (SEDs), and the Herschel PACS data present cold dusty cores which hints for new generations of star formation.

We present in Section~2 optical imaging of S155 as well as details of the retrieval of archival 2MASS and WISE data.  In Section~3, we discuss the spatial distribution of the sub-clusters toward S155. SED classification of the sample sources follows in Section~4. The results achieved are discussed in Section~5 and summarized in Section~6.


\section{Data acquisition and analysis}
\subsection{Narrowband Imaging}
S155 was observed with the MOSAIC camera on the Mayall 4-meter telescope at Kitt Peak National Observatory. MOSAIC is an optical camera. It consists of eight 2048$\times$4096 CCD detectors arranged to form a 8192$\times$8192 array with 35-50 pixel wide gaps between the CCDs. With a scale of $0\farcs26$ pixel$^{-1}$, the field of view is about $36\arcmin\times 36\arcmin$. To fill in the bad columns and gaps, all observations are completed in a five-exposure dither pattern with offsets of 100 pixels.

Observations toward S155 were obtained on 9 September 2009 with the ``Nearly-Mould" $I$ (MOSAIC filter k1005), H$\alpha$ (k1009) and [S$\textsc{ii}$] (ha16, H-alpha+16nm, k1013) filters, whose central wavelengths/FWHMs are 6574.74/80.62, 6730.72/81.1, and 8204.53/1914.59 \AA, respectively.  Five exposures of 300~sec were obtained in H$\alpha$ and [S$\textsc{ii}$]. And five 180~sec exposures were obtained in $I$.
\subsection{2MASS JHK$\textsc{s}$ photometry}
Infrared (IR) sources in the field of S155 were found using archival data from the 2MASS PSC. To ensure the reliability of the extracted sources, we employed the following strict requirements in the sample selection, which are revised based on the criteria presented by \cite{2005A&A...431..925L}: (1) $[JHK_S]_{-}cmsig \leq 0.1$ (corrected $JHK_S$ band photometric uncertainty less than or equal to 0.1 mag) (2) $K_{S-}snr > 15$ ($K_S$ band ``scan" signal-to-noise ratio greater than 15).
\subsection{WISE photometry}

The Wide-field Infrared Survey Explorer (WISE) is a NASA medium-class explorer mission, launched on 2009 December 14 \citep{2010AJ....140.1868W,2011ApJ...735..112J}. WISE mapped the entire sky simultaneously in four bands centered at 3.4, 4.6, 12, and 22 $\mu$m (W1, W2, W3, and W4) with $5\sigma$ point-source sensitivities of approximately 0.08, 0.1, 1, and 6 mJy \citep{2012ApJ...744..130K}.

In this paper, the sample of excessive emission sources selected based on the 2MASS data is cross-identified with the WISE All-Sky Source Catalog using the simple positional correlation method with a $3^{\prime\prime}$ search radius \citep{2012ApJ...744..130K}.
The following sample selection criterion was employed to guarantee the reliability of the WISE data in use and to allow a rigorous analysis. Only sources with certain detection in all four bands (W1 W2 W3$_{sigmpro}\leq 0.1$ and W4$_{sigmpro}\leq 0.5$) were considered.

\section{Clustering of star formation in S155}

 Figure~\ref{HIST} shows the H$\alpha$ image of a $30^\prime\ \times\ 30^\prime$ region surrounding S155. The bright portion delineated with a green line is the H~$\textsc{ii}$ region which is the interface between the Cep B molecular cloud and the association Cep OB3b. In this section, we will explore the 2MASS sources in the field shown in Figure~\ref{HIST}.

 \subsection{Color-color diagram}

In the target region, 2MASS database contains more than ten thousands photometric detections. We narrowed down the catalogue to 4481 sources using the select criteria mentioned in Section 2.

 \begin{figure}
   \centering
   \includegraphics[width=120mm]{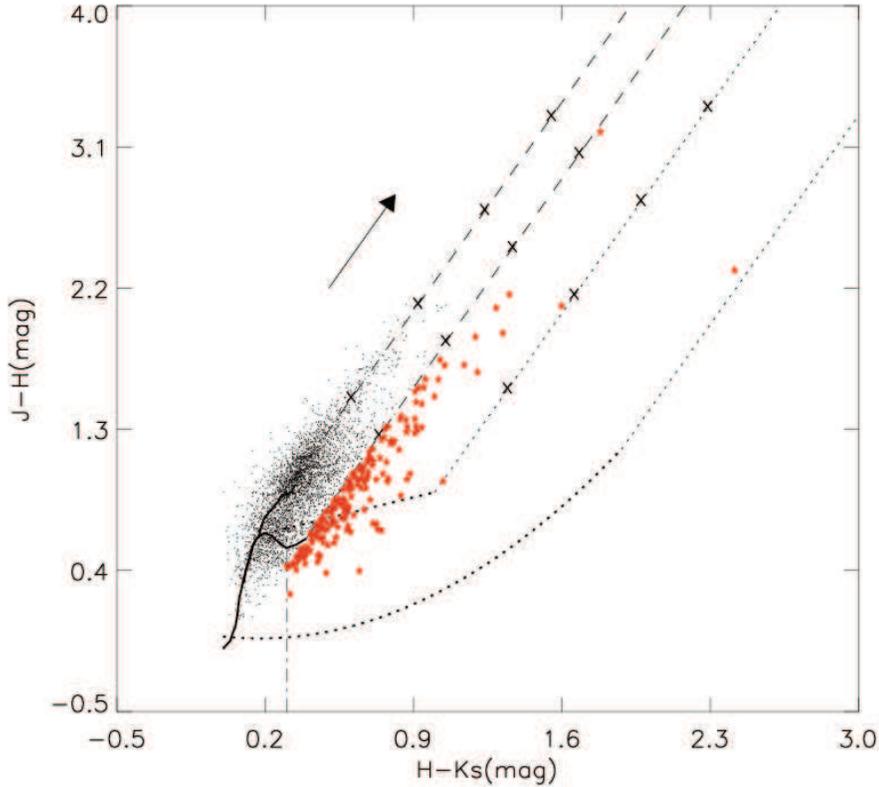}
   \begin{minipage}[]{100mm}
\caption{ $(J-H)$ vs. $(H-K_S)$ C-C diagram of S155. The sample sources are denoted as dots. Selected IR excessive emission sources are labeled in red. Black solid lines represent the loci of the MS dwarfs and giant stars \citep{1988PASP..100.1134B}. Two paralleled dashed lines define the reddening band for normal stars. The arrow shows a reddening vector of $A_v=5$ mag \citep{1985ApJ...288..618R}. The left dotted line indicates the locus of dereddened T~Tauri star \citep{1997AJ....114..288M} and its reddening band boundary. The right dotted line indicates the locus of dereddened HAeBe \citep{1992ApJ...393..278L} and its reddening direction. Crosses were over plotted with an interval corresponding to 5 mag of visual extinction.}
\label{CCD}
\end{minipage}
\end{figure}

2MASS $JHK_S$ color-color (C-C) diagram has been widely used to select sources with IR excessive emissions. Figure \ref{CCD} shows the C-C diagram for S155. All the 2MASS sources that match our criteria are put into the $JHK_S$ C-C diagram and denoted as dots. We defined 217 objects as YSO candidates for they are located below the right line of the normal star reddening band and $H-K_S > 0.3$ in the C-C diagram. Those sources are selected for they possess intrinsic color excesses likely originating from emission of circumstellar dust, commensurate with their embedded nature. The color distinction helps to eliminate foreground field stars and narrowed the YSO candidates sample.

\subsection{Surface density distribution}

   \begin{figure}
   \centering
   \includegraphics[width=11cm, angle=0]{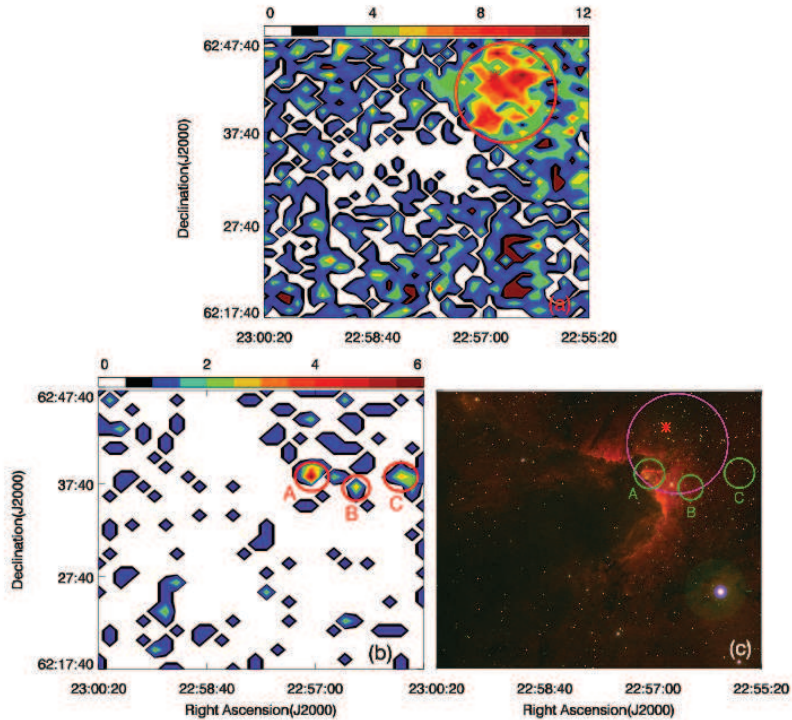}
   \begin{minipage}[]{140mm}
   \vspace{5mm}
  \caption{(a) Surface density distribution of the 2MASS sources toward S155. The densest region at the top right corner marked with a large red circle is consistent with the position of Cep OB3b, and the asterisk indicates HD~217086 (R.A.= $22^h56^m47^s.19$, decl.= $+62^\circ 43^{\prime}37^{\prime\prime}.64$(J2000)). The color bar indicates the number of sources in every 0.75$\times$0.75 arcmin$^2$. (b) Spatial distribution of the 217 2MASS excessive emission sources. Red circles mark the three densest regions of YSO candidates, which harbor active star formation. (c) Trichromatic image of S155 generated from I (blue), [S$\textsc{ii}$] (green), and H$\alpha$ (red) band image observed with the Mayall 4-meter telescope of KPNO. Red asterisk indicates the main exciting source HD~217086.}
\label{DMA} \end{minipage}
   \end{figure}

Shown in Figure~\ref{DMA} (a) is the surface density map of all the validly detected 2MASS sources in the targeted region. Irregular distribution of sources is prominently visualized. There is a cavity in the center and a enhancement in the top-right corner. This prominently dense region is labeled with a large red circle and having the exciting star HD~217086 enclosed. There are about 12 sources in 0.75$\times$0.75 arcmin$^2$ at the densest region, which coincides with the location of Cep OB3b. However, the structure of the H~$\textsc{ii}$ region is not revealed in this panel.

In Figure \ref{DMA} (b), the spatial distribution of the YSO candidates, there are three densest regions revealed. Combined with the trichromatic image from the Mayall 4-meter telescope of KPNO presented in panel (c), the three subclusters all surround Cep OB3b region, likely be triggered by the stellar winds from those OB stars. Compared to cluster B and C, subcluster A corresponds to the bright rim marked in panel (c), and is more compact and brighter, which may contains younger star formation activity.

\section{Identification and classification of excess emission sources in S155}

\subsection{Selection of YSOs based on WISE photometry}

\cite{2012ApJ...744..130K} have developed a scheme to identify YSOs based on WISE colors and magnitudes. With this scheme, contaminants can be removed and YSO candidates can be roughly classified. Contaminants arises from non-YSO sources, including star-forming galaxies, broad-line active galactic nuclei (AGNs), unresolved knots of shock emission from outflows colliding with cold cloud material, planetary nebulae, and asymptotic giant branch (AGB) stars \citep{2012ApJ...744..130K}. And YSOs can be classified into the canonical categories of Class I, Class II, with supplemental categories of ``deeply embedded sources"(added to the Class I tallies) and ``transition disks" \citep[objects with optically thick excess emission at long wavelengths and little to no excess at short wavelengths,][]{1989AJ.....97.1451S}. After cross-identified with WISE data, 55 sources remain in the YSO candidates sample. Based on the scheme, 11 contaminants were removed and 44 were classified as YSOs.

\subsection{Classification based on SED fitting}
To more accurately determine their evolutionary status, the SEDs of the 44 YSOs are fitted. \cite{2006ApJS..167..256R} developed a grid of 200,000 YSO models to fit the SED from optical to millimeter wavelengths. Those models span a wide range of evolutionary stages for different stellar masses. They also provide a linear regression tool, by which we can select all model SEDs that fit the observed SED better than a specified $\chi^2$ \citep{2006ApJS..167..256R}.

\cite{1987IAUS..115....1L} developed a widely used classification scheme for YSOs, based on the ``four staged" star formation scenario proposed by \cite{shu87}. With an evolutionary sequence from early type to late type, YSOs were classified into Class I to III, primarily based on their SEDs. \cite{2006ApJS..167..256R} presented a classification scheme refers to the actual evolutionary stage of the object based on physical properties. However, in view of the differences between observable and physical properties, we primarily refer the ages fitted by the tool and the slope of its near/mid-IR SED. Class I sources refer to objects whose $Age\approx10^5$ yr and $Slope_{near/mid-IR}> $ 0, Class II sources refer to $Age\approx10^6$ yr and $Slope_{near/mid-IR}\le $ 0, and Class III is for objects whose SED is similar to a black-body spectrum.

\begin{figure}
   \centering
   \includegraphics[width=140mm, angle=0]{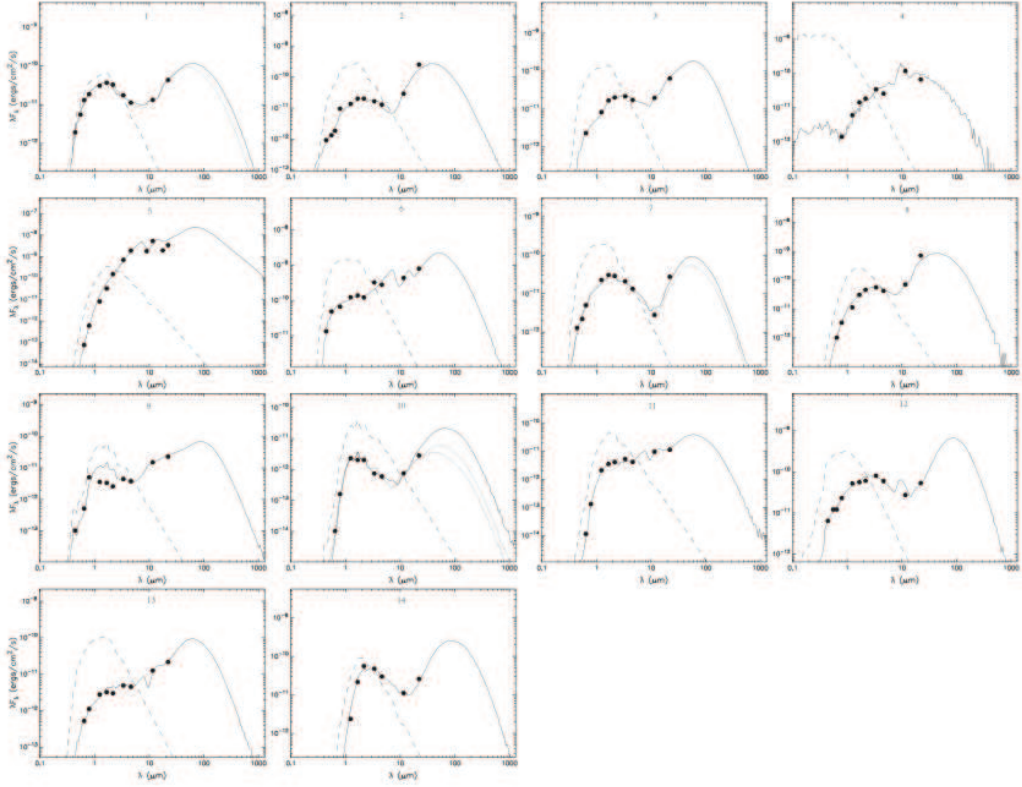}
   \begin{minipage}[]{140mm}
   \caption{Results of the SED fitting for the sample sources. The black dots show the measured fluxes currently available. The solid black curve represents the best model fitting of the data points with the smallest $\chi^2$. The gray curves indicate the other potential model fitting results with $\chi^2$/N-$\chi^2_{best}/N$$<3$. The dashed curve gives the photosphere that is used as input for the radiative transfer code.}
   \label{SED}
   \end{minipage}
   \end{figure}

Multi-wavelength online archived data are used for SED fitting. In addition to WISE and 2MASS PSC data, we used: 1) IRAC and MIPS photometry at 3.6, 4.5, 5.8, 8 and 24m for a few of sources; 2) the far-IR IRAS Point Sources photometry at 12, 25, 60 and 100$\mu m$; 3) the mid-IR A (8.28 $\mu m$), C (12.13 $\mu m$), D (14.65 $\mu m$) provided by the MSX6C IR PSC, and S9W (9 $\mu m$), L18W (18 $\mu m$) provided by the AKARI IRC PSC; 4) BVR photometry from the Naval Observatory Merged Astrometric Dataset (NOMAD). As a result, SEDs are fitted with a photometric catalogue over a large wavelength range. The PMS nature and evolutionary status of all the YSO candidates have been confirmed and corrected by results of the SED fittings (Table \ref{ALL}). Figure~\ref{SED} illustrates the SED of the CLASS I sources of the sample.

\section{Discussion}

\subsection{Sequential and triggered star formation around S155}

  \begin{figure}
   \centering
   \includegraphics[width=10cm, angle=0]{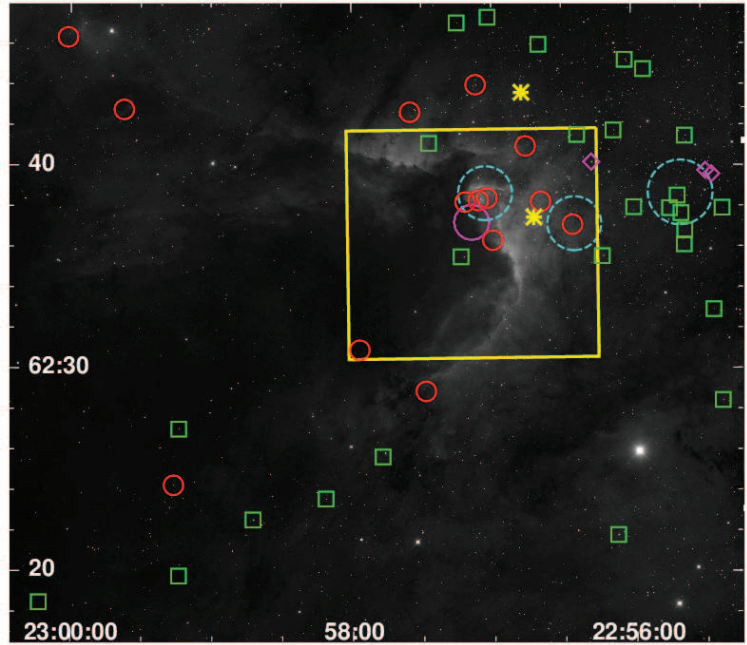}

   \begin{minipage}[]{100mm}

   \caption{Distribution of YSOs classified based on SEDs. CLASS I, II, and III sources are indicated with red circles, green boxes, and magenta diamonds, respectively. The yellow asterisks indicate HD~217086 (upper) and HD~217061. Cyan dashed circles present the three subclusters marked in Figure~\ref{DMA}. Magenta circle is the location of the dense core in Cep B molecular. The Herschel PACS image region is outlined in yellow.}
   \label{YSOD} \end{minipage}
   \end{figure}

In this work, we provide new additional evidence to support a scenario of sequential and triggered star formation in S155. Figure~\ref{YSOD} presents the H$\alpha$ image of S155, on which the sample sources classified based on SEDs are overlaid with different symbols. The spatial distribution of the classified sources reveals a dramatic spatiotemporal gradient: younger stars (Class I sources) are clustered in the H~$\textsc{ii}$ region S155, while older stars (Class II and Class III sources) are dispersed in the other side of the primary ionizing star HD~217086. Of the considered locations of three subclusters in Figure~\ref{DMA}, region A is indeed younger and more compact than the other two. Furthermore, most of the identified CLASS I and CLASS II sources are located along the edge of the molecular cloud. All these characters are consistent with a triggered nature of star and cluster formation in this region.

\subsection{New generation star formation}

   \begin{figure}
   \centering
   \includegraphics[width=14cm, angle=0]{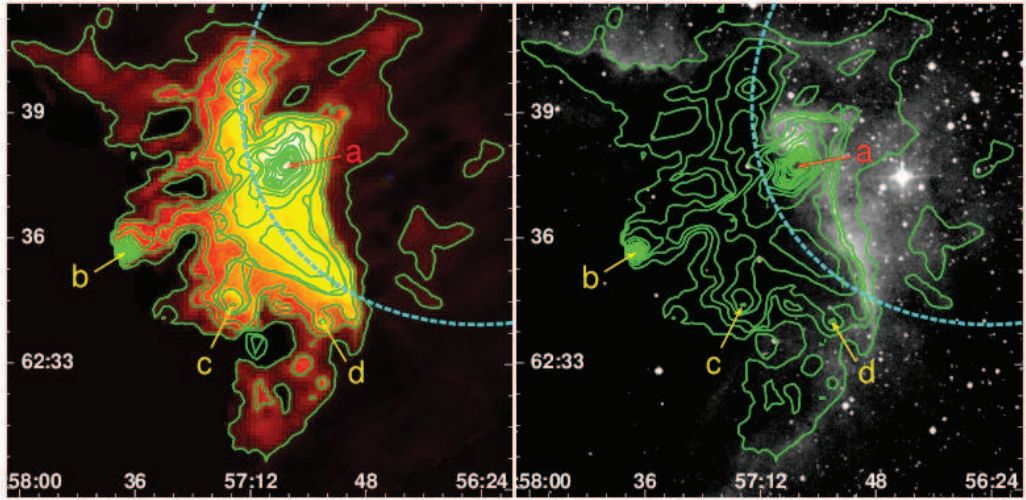}
   \begin{minipage}[]{140mm}

   \caption{Left: Color composite image of S155 and its nearby region in Herschel PACS 160 $\mu m$ (red), 70 $\mu m$ (green) and WISE W3 12 $\mu m$ (blue). Right: DSS 2 red band image of target region. Green contours in both two images are generated from the Herschel PACS image in 160 $\mu m$ band. }
   \label{PACS}
   \end{minipage}

   \end{figure}

 To further investigate the interstellar materials within the molecular cloud, Herschel PACS \citep{2010A&A...518L...2P} images of a region outlined in Figure~\ref{YSOD} were employed. The left panel of Figure~\ref{PACS} presents the composite image of this target region, which was compiled with the PACS 160 $\mu m$ (red), 70 $\mu m$ (green) and WISE W3 (blue) imaging data. It is evident that the dark edge of the molecular cloud in the DSS-2 red band image (left panel in Figure~\ref{PACS}) is bright in emission in the far-IR. This indicates the existence of large amount cold dust and star formation activity. The overlaid contours generated based on the Herschel PACS image indicate the presence of four far-IR cores. Among those bright cores, core ``a" is spatially consistent with the location of dense core in Cep B molecular cloud. Behind this dense core, an arc-shaped bright layer surround it and face to HD~217061. Its origin could be attributed to the compression from the feedback of HD~217061 to the molecular cloud. Based on distribution of them and other three cores, we infer that new generations of star formation is going on in the edge of molecular cloud, the morphology of the bright part is both effected by the illumination star HD~217061 and dense core.

 Based on these evidences, the scenario of the sequential and triggered star formation in the whole S155/Cep OB3b region is clear. Due to the influence of Cep OB3a, the first generation of the OB stars were born in Cep OB 3b, which contains HD~217086 and HD~217061. Then the wind originating from these stars compressed the surrounding cloud to trigger the second generation of stars (region A, B, and C marked in Figure~\ref{DMA}). On the east of Cep OB3b, a bright arc of nebulosity was produced for the existence of Cep molecular cloud, and defined the interface between the S155 H~$\textsc{ii}$ region and the Cep B molecular cloud \citep{2003MNRAS.341..805P}. As the surface of the cloud is being eroded, primarily by the illumination star HD~217061, the cloud edge moves eastward across the observer's field of view, with the third generation stars emerging from the obscuring molecular cloud (bright cores presented in Figure~\ref{PACS}).

\section{Summary}

We present a comprehensive study of sequential and triggered star formation toward the H~\textsc{ii} region S155. The sample of excessive emission sources are selected based on archived 2MASS data, which is then cross-identification with the WISE PSC. SED fittings are employed to further classify the IR sources. In the target region, we identify 14 Class~I, 27 Class~II and 3 Class~III sources, 20 of which are newly discovered.

The spatial distribution of the IR excessive emission sources
selected based on their 2MASS colors reveals the existence of three
sub-clusters in this region, which may have a triggered origin due
to their spatial distribution as respect to the massive cluster Cep
OB3b. Compared with the other two subclusters, region~A is spatially
associated with the brightest part of S155 and harbors the most
Class I sources, which supports the conclusion that A might be
younger than the other two regions.  This is supported by the
observation it is located on the interface between S155 and Cep B
molecular cloud. Based on the spatial distribution of the classified
YSOs, all Class I sources are found to located within S155, while
the Class II sources are distributed on the outer edges of the HII
region, which suggests a scenario of sequential and/or triggered
star formation in this region. Furthermore, four far-IR cores and a
bright arc were discovered by the Herschel PACS data, which provide
evidence for new generations of star formation.

\begin{deluxetable}{cccccccccccc}
\tabletypesize{\scriptsize} \rotate \tablecaption{Photometry and
classification of YSO sample sources.\label{ALL}} \tablewidth{0pt}
\tablehead{ \colhead{ID} & \colhead{R.A.(J2000.0)} &
\colhead{Dec.(J2000.0)} & \colhead{J} & \colhead{H} & \colhead{Ks} &
\colhead{W1} & \colhead{W2} & \colhead{W3} & \colhead{W4} &
\colhead{M$^d$} & \colhead{Class} \\
\colhead{} & \colhead{} & \colhead{} & \colhead{(mag)} & \colhead{(mag)} &
\colhead{(mag)} & \colhead{(mag)} & \colhead{(mag)} &
\colhead{(mag)} & \colhead{(mag)} &
\colhead{(M$_\odot$)} & \colhead{SEDs}\\
\colhead{(1)} & \colhead{(2)} & \colhead{(3)} & \colhead{(4)} & \colhead{(5)} &
\colhead{(6)} & \colhead{(7)} & \colhead{(8)} &
\colhead{(9)} & \colhead{(10)} &
\colhead{(11)} & \colhead{12}}
\startdata
 1$\tablenotemark{a}$&22 56 25.74&62 37 07.0&$12.739\pm0.021$&$11.754\pm0.022$&$11.122\pm0.02$&$10.493\pm0.026$&$9.993\pm0.023$&$6.973\pm0.019$&$3.539\pm0.037$&0.72&I\\
 2$\tablenotemark{b}$&22 56 39.30&62 38 15.0&$13.632\pm0.019$&$12.397\pm0.022$&$11.648\pm0.02$&$10.556\pm0.024$&$9.833\pm0.02$&$6.115\pm0.019$&$1.623\pm0.027$&1.16&I\\
 3$\tablenotemark{a}$&22 56 45.65&62 40 59.1&$14.166\pm0.03$&$12.599\pm0.021$&$11.651\pm0.018$&$10.265\pm0.027$&$9.514\pm0.022$&$6.552\pm0.039$&$3.117\pm0.064$&0.73&I\\
 4$\tablenotemark{a}$&22 56 59.85&62 36 20.8&$14.512\pm0.033$&$12.769\pm0.023$&$11.744\pm0.018$&$9.765\pm0.029$&$9.084\pm0.021$&$4.605\pm0.017$&$3.077\pm0.161$&0.8&I\\
 5$\tablenotemark{a}$&22 57 02.20&62 38 25.1&$14.182\pm0.052$&$11.868\pm0.039$&$9.453\pm0.019$&$6.48\pm0.024$&$4.426\pm0.022$&$0.473\pm0.022$&$-1.2\pm0.007$&3.01&I\\
 6$\tablenotemark{a}$&22 57 05.94&62 38 18.0&$11.244\pm0.02$&$10.316\pm0.021$&$9.713\pm0.016$&$7.329\pm0.01$&$6.514\pm0.009$&$3.201\pm0.075$&$0.385\pm0.023$&2.48&I\\
 7$\tablenotemark{a}$&22 57 06.87&62 44 00.8&$13.071\pm0.022$&$11.949\pm0.021$&$11.256\pm0.018$&$10.3\pm0.023$&$9.806\pm0.02$&$8.661\pm0.077$&$4.034\pm0.06$&0.73&I\\
 8$\tablenotemark{a,c}$&22 57 11.54&62 38 13.6&$13.827\pm0.026$&$11.937\pm0.02$&$10.745\pm0.016$&$9.22\pm0.033$&$8.537\pm0.032$&$5.146\pm0.095$&$0.497\pm0.025$&2.08&I\\
 9&22 57 28.67&62 28 54.1&$15.08\pm0.052$&$14.348\pm0.051$&$13.875\pm0.063$&$11.98\pm0.024$&$11.177\pm0.022$&$6.837\pm0.018$&$4.25\pm0.026$&&I\\
10&22 57 35.03&62 42 39.9&$15.559\pm0.067$&$14.903\pm0.076$&$14.154\pm0.066$&$13.929\pm0.033$&$13.174\pm0.034$&$10.089\pm0.06$&$6.502\pm0.081$&&I\\
11&22 57 56.81&62 30 56.2&$15.637\pm0.069$&$14.27\pm0.053$&$13.364\pm0.042$&$11.779\pm0.029$&$11.043\pm0.035$&$7.302\pm0.086$&$4.967\pm0.066$&&I\\
12&22 59 16.17&62 24 15.4&$12.171\pm0.021$&$11.293\pm0.031$&$10.453\pm0.021$&$8.836\pm0.023$&$8.177\pm0.02$&$6.184\pm0.034$&$3.313\pm0.038$&&I\\
13&23 00 01.24&62 46 19.9&$15.351\pm0.061$&$14.388\pm0.071$&$13.718\pm0.055$&$11.884\pm0.03$&$10.977\pm0.031$&$7.042\pm0.018$&$4.299\pm0.031$&&I\\
14&22 59 37.21&62 42 46.3&$15.523\pm0.067$&$12.327\pm0.032$&$10.545\pm0.02$&$9.4\pm0.022$&$8.932\pm0.019$&$7.167\pm0.029$&$4.097\pm0.026$&&I\\
15&22 55 22.49&62 28 25.1&$14.19\pm0.026$&$13.072\pm0.025$&$12.331\pm0.023$&$11.064\pm0.022$&$10.494\pm0.021$&$8.215\pm0.057$&$6.307\pm0.08$&&II\\
16&22 55 38.15&62 36 06.3&$15.055\pm0.037$&$13.763\pm0.037$&$12.845\pm0.023$&$11.636\pm0.023$&$10.986\pm0.02$&$9.105\pm0.032$&$7.077\pm0.094$&&II\\
17$\tablenotemark{a}$&22 55 55.00&62 44 45.8&$15.558\pm0.072$&$13.994\pm0.039$&$13.071\pm0.028$&$11.685\pm0.024$&$10.947\pm0.021$&$8.574\pm0.028$&$6.336\pm0.063$&0.25&II\\
18$\tablenotemark{a}$&22 56 23.57&62 41 32.6&$13.377\pm0.025$&$12.286\pm0.028$&$11.606\pm0.023$&$10.433\pm0.026$&$9.916\pm0.02$&$7.622\pm0.022$&$4.803\pm0.035$&0.53&II\\
19&22 55 21.78&62 37 53.5&$12.693\pm0.019$&$11.381\pm0.021$&$10.442\pm0.018$&$9.427\pm0.023$&$8.575\pm0.019$&$6.197\pm0.016$&$4.266\pm0.024$&&II\\
20&22 55 25.89&62 32 53.8&$12.546\pm0.019$&$11.599\pm0.02$&$10.962\pm0.014$&$10.174\pm0.023$&$9.626\pm0.021$&$7.224\pm0.018$&$5.072\pm0.03$&&II\\
21$\tablenotemark{a}$&22 55 37.45&62 41 28.0&$12.567\pm0.027$&$11.249\pm0.023$&$10.468\pm0.018$&$9.129\pm0.023$&$8.399\pm0.018$&$6.546\pm0.012$&$4.828\pm0.028$&1.81&II\\
22&22 55 37.88&62 36 47.5&$15.666\pm0.066$&$14.362\pm0.053$&$13.597\pm0.039$&$12.599\pm0.025$&$11.958\pm0.023$&$10.416\pm0.068$&$7.77\pm0.185$&&II\\
23$\tablenotemark{b}$&22 55 39.40&62 37 39.0&$14.077\pm0.03$&$12.97\pm0.032$&$12.286\pm0.022$&$11.293\pm0.023$&$10.676\pm0.021$&$9.058\pm0.03$&$6.696\pm0.065$&0.61&II\\
24$\tablenotemark{b}$&22 55 40.87&62 38 30.7&$15.758\pm0.072$&$14.584\pm0.067$&$13.876\pm0.052$&$12.829\pm0.026$&$12.143\pm0.025$&$10.351\pm0.057$&$7.986\pm0.232$&0.24&II\\
25$\tablenotemark{b}$&22 55 44.17&62 37 53.8&$15.208\pm0.05$&$14.239\pm0.052$&$13.463\pm0.034$&$12.424\pm0.025$&$11.737\pm0.023$&$10.348\pm0.06$&$7.507\pm0.135$&0.25&II\\
26$\tablenotemark{b}$&22 55 59.56&62 37 57.2&$14.818\pm0.039$&$13.946\pm0.039$&$13.34\pm0.039$&$12.581\pm0.026$&$12.051\pm0.024$&$10.195\pm0.057$&$6.517\pm0.07$&0.28&II\\
27$\tablenotemark{a}$&22 56 02.81&62 45 13.7&$13.338\pm0.024$&$12.282\pm0.023$&$11.631\pm0.02$&$10.705\pm0.023$&$10.09\pm0.02$&$8.066\pm0.026$&$6.097\pm0.066$&0.51&II\\
28&22 56 07.67&62 21 48.2&$10.44\pm0.024$&$9.47\pm0.024$&$8.431\pm0.017$&$6.931\pm0.034$&$6.09\pm0.024$&$2.989\pm0.014$&$1.149\pm0.013$&&II\\
29$\tablenotemark{a}$&22 56 07.98&62 41 45.1&$13.069\pm0.019$&$11.86\pm0.019$&$11.064\pm0.018$&$10.107\pm0.023$&$9.351\pm0.02$&$7.274\pm0.02$&$5.313\pm0.036$&0.93&II\\
30$\tablenotemark{a}$&22 56 13.18&62 35 33.1&$11.805\pm0.019$&$10.781\pm0.019$&$10.122\pm0.017$&$8.994\pm0.022$&$8.47\pm0.02$&$6.576\pm0.017$&$3.987\pm0.028$&1.88&II\\
31&...&...&...&...&...&...&...&...&...&...&...\\

\enddata
\tablenotetext{*}{Table \ref{ALL} is published in its entirety in the electronic edition. A portion is shown here for guidance regarding its form and content.}
\tablenotetext{a}{Sources detected by \cite{2006ApJS..163..306G}.}
\tablenotetext{b}{Sources detected by \cite{2009ApJ...699.1454G}.}
\tablenotetext{c}{Sources detected by \cite{1993A&A...273..619M}.}
\tablenotetext{d}{From the results of \cite{2006ApJS..163..306G} and \cite{2009ApJ...699.1454G}.}
\end{deluxetable}

\normalem
\begin{acknowledgements}
We appreciate very much the helpful comments and suggestions from the referee.
This work employed data from 2MASS, WISE, Herschel and other astronomical databases. Our investigation is supported by funding from the National Natural Science Foundation of China (Grant Nos. 11073027 and 11003021), Beijing Natural Science Foundation (1144015) and the Department of International Cooperation of the Ministry of Science and Technology of China (Grant No. 2010DFA02710). The Mayall 4-meter telescope at Kitt Peak National Observatory is a facility of the National Optical Astronomy Observatory, which is operated by the Association of Universities for Research in Astronomy, Inc. (AURA) under cooperative agreement with the National Science Foundation.
\end{acknowledgements}
\clearpage


\label{lastpage}

\end{document}